\documentclass[draft,sw]{agutex2015}

\usepackage{url}
\usepackage[table,xcdraw]{xcolor}
 \usepackage{lineno}
\usepackage{pdflscape}
 \usepackage[dvips]{graphicx}
\setkeys{Gin}{draft=false}
\authorrunninghead{BHASKAR AND Vichare.}

\titlerunninghead{Prediction of SYMH and ASYH indices}

\authoraddr{Corresponding author: Ankush Bhaskar,
Indian Institute of Geomagnetism, New Panvel, Navi Mumbai, 410218,  India.
(ankushbhaskar@gmail.com)}

\begin{document}

%
%

\title{Prediction of SYMH and ASYH indices for geomagnetic storms of solar cycle 24 including recent St. Patrick's day, 2015 storm using NARX neural network}
%
%

%
%



 \authors{Ankush  Bhaskar \altaffilmark{1}
 and  Geeta Vichare \altaffilmark{1}}

\altaffiltext{1}{Indian Institute of Geomagnetism, New Panvel, Navi Mumbai, 410218,  India.}





%
%


\keypoints{ \item NARX network is developed for prediction of SYMH index  
\item First ANN-based network for ASYH index.
\item Prediction of SYMH and ASYH during St. Patrick's day, 2015 geomagnetic storm }


%
%


\begin{abstract}
Artificial Neural Network (ANN) has proven to be very successful in forecasting variety of irregular magnetospheric/ionospheric processes like geomagnetic storms and substorms. SYMH and ASYH indices represent longitudinal symmetric and asymmetric component of the ring current. Here, an attempt is made to develop a prediction model for these indices using ANN. The ring current state depends on its past conditions therefore, it is necessary to consider its history for prediction. To account this effect \textit{Nonlinear Autoregressive Network with eXogenous inputs} (NARX) is implemented. This network considers input history of 30 minutes and output feedback of 120 minutes. Solar wind parameters mainly velocity, density and interplanetary magnetic field are used as inputs. SYMH and ASYH indices during geomagnetic storms of 1998-2013, having minimum SYMH $<-85$ nT are used as the target for training two independent networks. We present the prediction of SYMH and ASYH indices during 9 geomagnetic storms of solar cycle 24 including the recent largest storm occurred on  St. Patrick's day, 2015. The present prediction model reproduces the entire time profile of SYMH and ASYH indices along with small variations of $\sim 10-30 minutes$ to good extent within noise level, indicating significant contribution of interplanetary sources and past state of the magnetosphere. Therefore, the developed networks can predict SYMH and ASYH indices about an hour before, provided, real-time upstream solar wind data is available. However, during the main phase of major storms, residuals (observed-modeled) are found to be large, suggesting influence of internal factors such as magnetospheric processes. 
\end{abstract}

%
%

%

\begin{article}-

%
%

\section{Introduction}

Transient ejections from the Sun set up large scale disturbances in the interplanetary  space. These disturbances interact with the Earth's magnetic field, resulting into the severe space weather events, such as geomagnetic storm, substorm etc. As the present space-technology is vulnerable to the geomagnetic disturbances, predicting geomagnetic field response well in advance is an important aspect of space weather studies. Long duration southward interplanetary magnetic field injects solar wind energy into the Earth's magnetosphere-ionosphere system mainly through reconnection\citep{gonzalez1994geomagnetic}. This results in the azimuthal drift of the charged particles inside the magnetosphere, establishing ring current in the equatorial plane. Intensification (main phase) and decay (recovery) of the storm time ring current consist of different processes. The main phase is primarily controlled by the solar wind conditions, whereas decay of the ring current has a major contribution from the internal magnetospheric processes. Due to varying nature of the storm sources, the magnetospheric dynamics and the energy budget  involved in each storm differs considerably\citep{Vichare2005}. The injection of solar wind particles and transmission of solar wind electric field generate various currents in the magnetosphere-ionosphere system such as cross-tail current, field aligned currents, partial ring current etc\citep{ohtani2000magnetospheric}. Moreover, sudden variations in the dynamic pressure of the solar wind alters magnetopause current and tail current. Also, they produce transient ionospheric currents\citep{vichare2014ionospheric}. The recovery phase of the ring current during geomagnetic storm has influence of various nonlinear phenomenon like wave-particle interaction, charge exchange, ionospheric outflow of $O^{+}$ ions, particle precipitation etc\citep{Daglis1999}. Superposed effect of these currents and magnetospheric nonlinear processes in the magnetosphere-ionosphere system makes prediction of storm-time temporal variations of ring current a challenging task.  

Ground magnetometer measures integrated effect of all these disturbed time and also quiet time ionospheric and magnetospheric currents. Geomagnetic indices like Disturbance storm time index (Dst) and Symmetric H-component (SYMH) index mainly represent ring current intensity during geomagnetic storms \citep{sugiura1964hourly, rangarajan1989indices, wanliss2006high}, derived using longitudinally distributed chain of low latitude ground-based magnetometers. SYMH is same as Dst, but it has 1 minute temporal resolution, which is very useful to study short temporal variations during the geomagnetic disturbances. SYMH is derived by first subtracting main geomagnetic field due to internal geodynamo and external Sq induced geomagnetic field variations and then averaging residual fields. Therefore, it is a good proxy for longitudinally symmetric component of the ring current. By removing globally symmetric component of the magnetic field variations from geomagnetic field variations at each station, longitudinally asymmetric geomagnetic field variations are derived. The range between maximum and minimum of these subtracted fields are compiled as ASYH index. ASYH have a significant contribution from various transient currents flowing in the magnetosphere-ionosphere system  such as currents associated with sudden impulses, solar flares, substorms and prompt penetration electric fields, partial ring current, field aligned currents, magnetotail current  etc \citep{clauer1980relative, iyemori1996decay,singh2013effect,singh2012solar}. Normally, during geomagnetic storms these asymmetric currents also get enhanced. Therefore, ASYH index is a good proxy for globally asymmetric currents in the magnetosphere-ionosphere system during geomagnetic storms. The contribution of substorms in ring current is a widely debated topic as some researchers believe to have significant contribution and some believe it is weak. \citep{newell2012supermag} showed that substorm affect in ring current is very small. Moreover, \citet{munsami2000storm} showed that when Dst station lies under a substorm current wedge, then only they observed significant contamination of ring current due to substorms. 

There are lot of efforts to understand the relationship between ring current (SYMH) and partial ring current (ASYH) respectively. Generally, it is observed that during the main phase of geomagnetic storm, ring current is highly asymmetric and becomes symmetric in the late recovery phase \citep{siscoe2012problem, jordanova2003ring}. \citet{liemohn2001dominant} reported that major part of magnetic field variations during the main phase of geomagnetic storms is due to asymmetric ring current. However, there are storms which show symmetric nature of the ring current even during the main phase which remains unexplained \citep{newell2012supermag}. The well known Love-Gannon relationship states that the difference between dawn and dusk disturbance-field (similar to ASHY index) at low latitudes is linearly proportional to Dst. However, \citep{siscoe2012problem} pointed out that this relationship can be explained only through field aligned currents.


As there are number of studies investigating relationship between symmetric and asymmetric ring current, at the same time efforts are going on to give more accurate prediction of these indices during geomagnetic storms. To forecast these geomagnetic indices (Dst, SYMH, AE etc) different approaches have been attempted \citep{williscroft1996neural, wu1996prediction, wu1998neural,weigel1999forecasting,o2000forecasting,wei2004prediction,gholipour2004black,boynton2011data,rastatter2013geospace,revallo2014neural,uwamahoro2014empirical}. These methods are mainly based on empirical or analytical relationships between solar wind and geomagnetic parameters, correlation and artificial neural networks (ANNs). Linear regression, statistical correlation etc have been proved to be useful in understanding storm time geomagnetic field variations. There are many empirical models for Dst prediction. A simple prediction algorithm for Dst index was proposed by \citet{burton1975empirical}, solely from a knowledge of the solar wind parameters. They assumed a constant ring current recovery time constant (e-folding time) for all the storms which may not be always a valid assumption. \citet{iyemori1980} first time successfully applied linear prediction filtering method for predicting geomagnetic activity using solar wind parameters. 

Artificial neural networks are being extensively used in many areas where nonlinear complexities are involved \citep{lippmann1987introduction,miller1993review,gardner1998artificial,unnikrishnan2014prediction}. In last few decades artificial neural networks are used for predicting geomagnetic activity at high and low geomagnetic latitude regions \citep{gleisner1997response,wu1996prediction}. There are many studies which attempted to predict symmetric part of the ring current and geomagnetic field variations using neural networks
 \citep{kamide1986solar,lundstedt1994prediction,wu1998neural,kugblenu1999prediction, unnikrishnan2012prediction,unnikrishnan2014prediction}. \citet{lundstedt1994prediction} developed feed forward neural network to predict geomagnetic activity index, Dst, one hour in advance. They could predict initial and main phase very well but the recovery phase was not modeled correctly. The feedforward networks do not have feedback from output or hidden nodes (refer section 2 for network architecture)  which constraint them in accurately modeling time series having memory. \citet{gleisner1996predicting} have used time-delayed feed-forward neural network for predicting Dst index. Further implementation of dynamic neural networks (i.e feedback networks) has improved the prediction accuracy for the recovery phase. Elman recurrent neural networks are implemented  by \citet{wu1997geomagnetic} to predict Dst during geomagnetic storms. The recovery part of the storm has modeled significantly well by Elman network as it takes feedback from the hidden layer. The predictions are generally very good for the main phase of the geomagnetic storms, but are fairly good for the recovery phases. The use of NARX network by \citet{cai2009prediction} for predicting SYMH has been very successful as it is observed to be better in performance due to feedback given from output node. However, the recent study by \citet{revallo2015modeling} developed prediction model for Dst using neural network and analytical prediction is done for two classes of geomagnetic storms caused by Coronal Mass ejection (CME) and Corotating Interaction Region (CIR). They observed better predication for CME driven storms than for CIR driven storm. 

ASYH is very valuable index to study the asymmetric development of magnetospheric storms during crossing of interplanetary disturbances (for example \citet{huttunen2006asymmetric}). There are number of studies trying to understand physical mechanism underlying the observed asymmetry in the ring current and its origin along with contribution of various currents in asymmetric ring current (e.g \citet{liemohn2001dominant, jordanova2003ring, newell2012supermag}). Though ASYH index is equally important during storm time dynamics, there are no reports of ANN based model available for ASYH index till date. The ANN based prediction of asymmetric ring current will help to understand the contribution of external/internal drivers in the observed asymmetry. Also, early forecast of ASYH will help space weather community to have a prior knowledge of the degree of asymmetry in the ring current. Therefore, developing ANN based prediction model for ASYH is the main objective of the present study. The present study develops NARX network based model to forecast both SYMH and ASYH indices, using interplanetary parameters as inputs and feedback from the output. For this purpose, we have used interplanetary parameters, SYMH and ASYH indices during major geomagnetic storms occurred between 1998-2015 covering around two solar cycles. We present prediction of SYMH and ASYH indices for the recent geomagnetic storm that took place on St. Patrick's day, 2015 (intense geomagnetic storm  of current solar cycle, 24) along with few other major storms from solar cycle, 24.

The paper is arranged as follows: Section 2 describes NARX neural network. Section 3 and 4 introduces data and training methodology. Section 5 discusses the network performance and prediction of SYMH and ASYH indices during geomagnetic storms. Paper ends with the discussion and conclusions in section 6.


\section{NARX Neural Network}

Artificial Neural Network (ANN) functions like biological neural network \citep{poulton2002neural}. The biological neuron is composed of dendrites, the soma and the axon. The neuron receives input signal from other neurons which are connected to its dendrites by synapses. The soma is mainly processing unit where inputs are integrated over space and time and it activates an output depending on the total input. This output is transmitted by the axon and distributed to other neurons by the synapses at the tree structure at the end of the axon \citep{herault1994reseaux}. The mathematical neuron functions little simpler way since integration takes place only over space. The inputs are given at one or many nodes called input nodes. The sum of these weighted inputs is performed at summing node which is fed to the nonlinear transform function or called as activation function to rescale the sum. The array of many nodes makes a network which can be made to learn relationships between inputs and targets, used for prediction.

\begin{figure}
\begin{center}
\noindent\includegraphics[width=12cm]{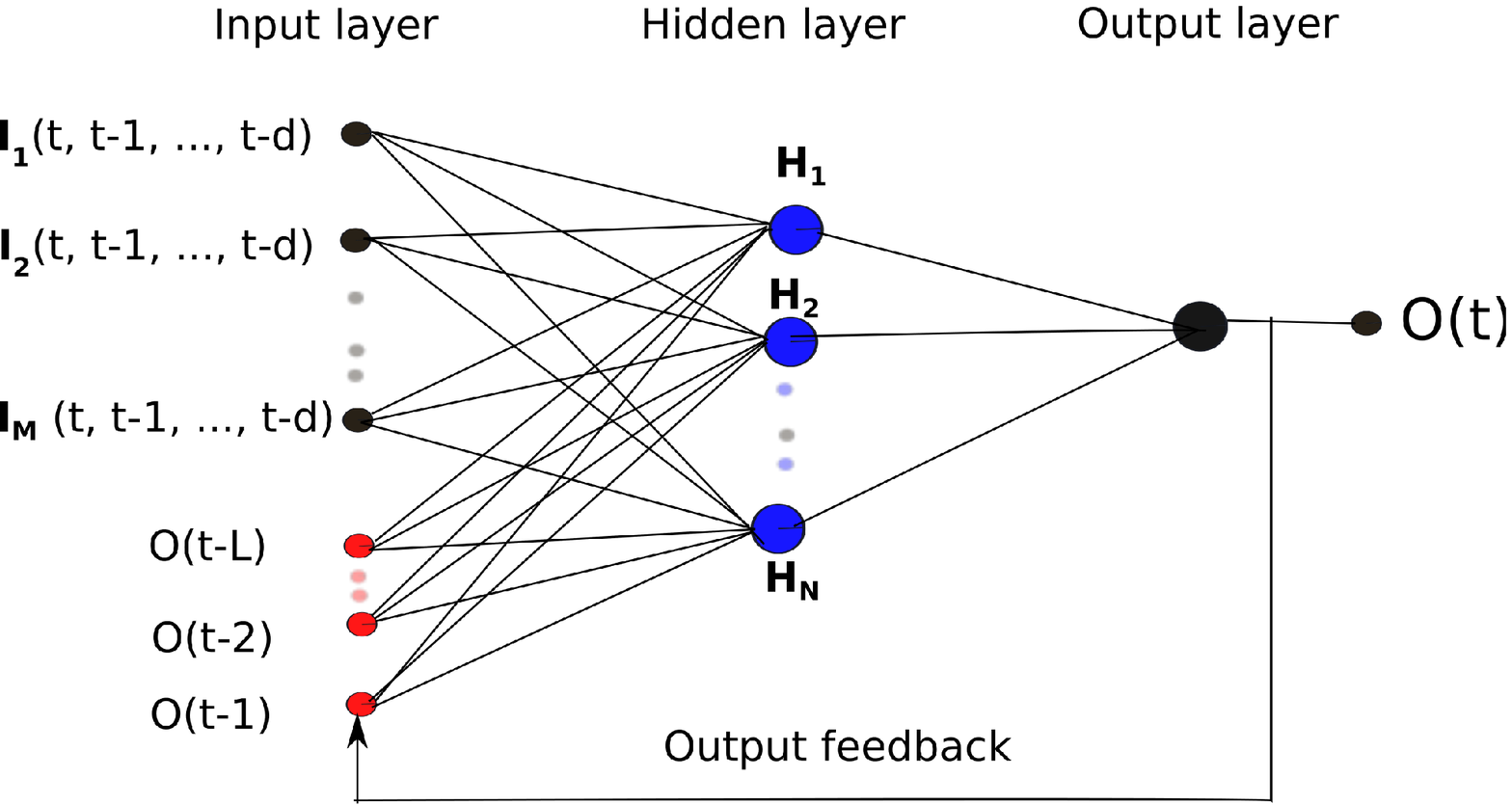}
\end{center}
\caption{Architecture of Nonlinear Auto Regressive with eXogenous inputs (NARX) network. d is input history length and L is output feedback length.}
\label{fig:narxnet}
\end{figure}

For the present study we have selected Nonlinear Auto Regressive with eXogenous inputs (NARX) model network due to its proven ability to account for the history of input and output parameters for prediction. This is feedback two-layer back propagation network with time-delayed feedback. The basic network architecture is presented in Figure~\ref{fig:narxnet}. As shown in the figure, inputs are shown to the network as temporal sequence with different time lags with time delay length, d. Whereas, past outputs of the network are provided to the NARX network as feedback having history length, L, which are called as context inputs. From left to right network has input layer, hidden layer and then output layer. The first layer or called as input layer receives externally provided values of input parameters. The second layer of the network does not see or act upon the external conditions hence termed as hidden layer. The hidden layer transforms the inputs such that the transformed inputs can be used by output layers. The output layer scales the hidden layer outputs to match the target. The dynamic behavior of the network can be formulated as 
\begin{equation}
O_{t} = F(O_{t-1},....O_{t-L}; I_{t},....I_{t-d})  
\end{equation}
Where, $O$ is the output of the network, $I$ denotes the input vector. Thus, the output of the NARX network is a function of present inputs and their past values along with history of the output. The inputs are processed by hidden nodes in the hidden layer, the output of $j^{th}$ hidden node is given by
\begin{equation}
H_{j}= tanh\left( \sum\limits_{i=1}^{M} W_{ji} I_{i}+\sum\limits_{l=1}^{L} W_{jl} C_{l} +b_{j} \right) 
\end{equation}
where $I_{i}$ is the value of input node i, M is total number of input nodes. $W_{ji}$ is a connecting weight of input node (i) and hidden node (j). Note that $tanh$ (hyperbolic tangent) is the transfer function for nodes in the hidden-layer. $b_{j}$ is bias of the $j^{th}$ neuron in hidden layer. Complex and nonlinear relationships between inputs and output are taken care by $tanh$ function. The output of the network ($O(t)$) is a linear summation over all hidden neuron outputs and output bias ($b_{0}$) which is represented by
\begin{equation}
O(t) =  \sum\limits_{j=1}^{s} W_{oj} H_{j}+ b_{0} 
\end{equation}
Here, $W_{oj}$ is connecting weight of hidden node to the output node.

\section{Database}
Different types of forecast models were studied prior to deciding the input and output database for NARX network. Interplanetary magnetic field, solar wind density and velocity are most crucial parameters controlling the storm profile. Also, the history has significant influence on the prediction accuracy. Therefore, we considered total interplanetary magnetic field (B) and its components (By and Bz), solar wind density (Nsw) and solar wind speed (Vsw) as input parameters. SYMH and ASYH indices are considered as target for the training two independent networks. 

The study is carried out considering 67 major geomagnetic storms (minimum SYMH $<-85$ nT) during 1998-2005 (Storm list is adopted from: \citet{cai2009prediction}) and 34 geomagnetic storms (minimum SYMH $<-85$ nT) identified between 2006-2015 (listed in Table \ref{table:tab1}). This period (1998-2015) covers $23^{rd}$ and ongoing $24^{th} $solar cycles having total 101 geomagnetic storms of minimum SYMH $<-85$ nT. It also includes 17th March geomagnetic storm which is a major storm of the $24^{th}$ solar cycle till date. Total 92 storms (1998-2013) are used for training and 9 storms occurred during 2104-15 are used to predict SYMH and ASYH indices. The utilized Solar wind parameters and indices were acquired from CDAWEB database (\url{http://cdaweb.gsfc.nasa.gov/}). One minute time resolution data was converted to five minute resolution for reducing the computation time. The missing data was interpolated using piecewise cubic hermite polynomial. Total data length of $\sim  685$ days having 5 minute resolution was used for developing the network. 

The data of 92 geomagnetic storms between 1998-2013 is used for learning the network, which is divided into the three  parts: (1)training ($75\%$), (2) validation ($15\%$) and (3) test($10\%$). As stated earlier the training data is used to learn the relationship between inputs and output. The validation of the network is determined through the identification of minimum error using $15\%$ of the data. Validation data is used to stop network from over-fitting the target. The test data was used to evaluate the performance of the network. Further, to check the prediction performance of the networks  9 geomagnetic storms are used, which occurred during January, 2014-July, 2015.

\section{Training}

ANN-based prediction model consists of mainly three steps:  training, validation and prediction \citep{haykin2004comprehensive}. 
The present network consist of one input layer with 30 external input nodes and 24 context inputs from the output, one hidden layer with 16 neurons and 1 output node. It is reported by earlier researchers that ring current history of about 2 hour is adequate for predicting SYMH index \citep{cai2009prediction}.  Also, communication time of interplanetary electric field from the Bow-shock nose to the equatorial ionosphere is observed to be $\sim 20$ minute \citep{bhaskar2013characteristics}. Therefore, the input history of 30 minute and output feedback length of 120 minute having 5 minute temporal resolution is used in the network. There are total 54 input nodes exist for each network. Input history of 30 minute implies the need of total 30 external input nodes for 5 input parameters (B, By, Bz, Nsw and Vsw ) each parameter having 6 nodes. Similarly, 120 minute feedback length from output makes 24 context inputs. 

The network is presented with the inputs to produce the desired output. For training the network, we have used a most popular \textit{back-propagation algorithm} \citep{rumelhart1985learning}. In this algorithm the weights are updated by using delta rule which is given by
\begin{equation}
 \Delta w(i+1)=-\eta \frac{dE}{dw}+\alpha.\Delta w(i)
\end{equation}

Here, $w$ represents weight of the nodes, $i$ is epoch, $\alpha$ and $\eta$ denote the momentum parameter and learning rate respectively. Momentum parameter is used to avoid local minimum, whereas learning rate controls the learning speed of the network. The $\alpha$ ranges between 0 to 1. For optimization of speed of learning the $\eta$ is adjusted in each iteration according to the performance of the network. For initialization, small random values are assigned to the network weights. Initially  $\alpha=0.9$ and $\eta=0.01$ were considered for the networks training.  $E$ is network error which is estimated by using the following equation which is also known as a cost function

\begin{equation}
 E=\frac{1}{2}\sum\limits_{k=1}^{N}\left( O^{k}-T^{k}\right)^2 
\end{equation}

$O$ is output of the network, $T$ is the target value and $N$ is the total number of training samples. The error is minimized during epoch to epoch of the training for obtaining final trained network. To avoid the over-fitting the validation dataset is used. During the training, error on the validation data is continually monitored. The training is terminated when the validation error reaches a minimum and then increases for next 6 epochs consecutively. It is known and observed that initialization of network weights and number of nodes in the hidden layer affect the performance of neural networks.  Hence, we trained the network multiple times by changing the initial weights and number of hidden layer nodes and selected the one which gave best results for prediction. 

Further, Root Mean Square Error ($RMSE$) was estimated to evaluate the performance of the network on test data consisting 9 geomagnetic storms occurred during January, 2014-July, 2015 including the recent geomagnetic storm of March 17, 2015 (see Table 1). The root mean square error can be computed as
\begin{equation}
RMSE = {\left[ {\frac{1}{N}\sum\limits_{{i = 1}}^N {{{(O^{i} - T^{i})}^2}} } \right]^{{1/2}}} 
\end{equation}

Also, the cross correlation coefficient (R) was estimated using equation (7) to quantify the similarities between time series of the observed and predicated SYMH/ASYH index.

\begin{equation}
 R = \frac{\sum\limits_{i=1}^{N} (T^{i}-\bar {T}) (O^{i}-\bar {O}) } { \sqrt{ (\sum\limits_{i=1}^{N} (T^{i}-\bar {T})^{2} } \sqrt{ (\sum\limits_{i=1}^{N} (O^{i}-\bar {O}) }  }
\end{equation}


\section{Results}
\subsection{Network performance }
\begin{figure}
\begin{center}
\noindent\includegraphics[width=14cm,angle=0]{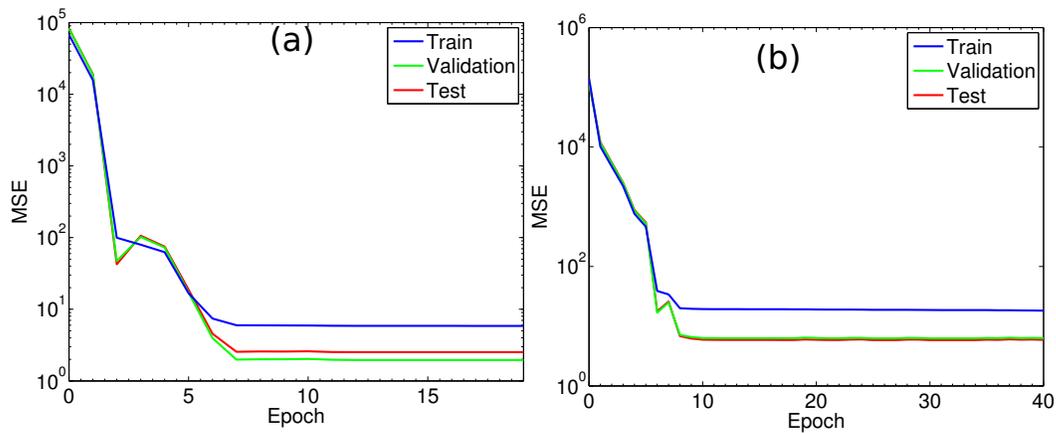}
\end{center}
\caption{The training performance of the neural network is shown for both (a) SYMH and (b) ASYH indices}
\label{fig:trainper}
\end{figure}

Figure ~\ref{fig:trainper} shows the performance of the trained networks. Figure ~\ref{fig:trainper}a shows the performance for SYMH whereas, Figure ~\ref{fig:trainper}b shows performance for ASYH index. The figure presents performance of all the steps i.e training, validation and the test. As a part of learning of the network, after each iteration the mean squared error of both the networks initially decreases. This characteristic is observed in both the panels of Figure ~\ref{fig:trainper} i.e initially the error in estimated SYMH and ASYH decreases with increasing epochs in similar fashion and then remain steady during training, validation and testing. One more common feature observed during training, validation and testing of SYMH and ASYH is that the errors of testing, and validation converge to a smaller value compared to the training. The best test performance of the SYMH network is achieved at Mean Squared Error (MSE), $\sim 6$ nT and that for ASYH network is  $\sim 18$ nT. This implies the training of SYMH network is better compared to ASYH. Note that, to prevent the network from over-fitting, the training was stopped when validation error increased continuously for the next six iterations. This is achieved at epoch=20 and 40 for SYMH and ASYH networks respectively.

\begin{figure}
\begin{center}
\noindent\includegraphics[width=14cm,angle=0]{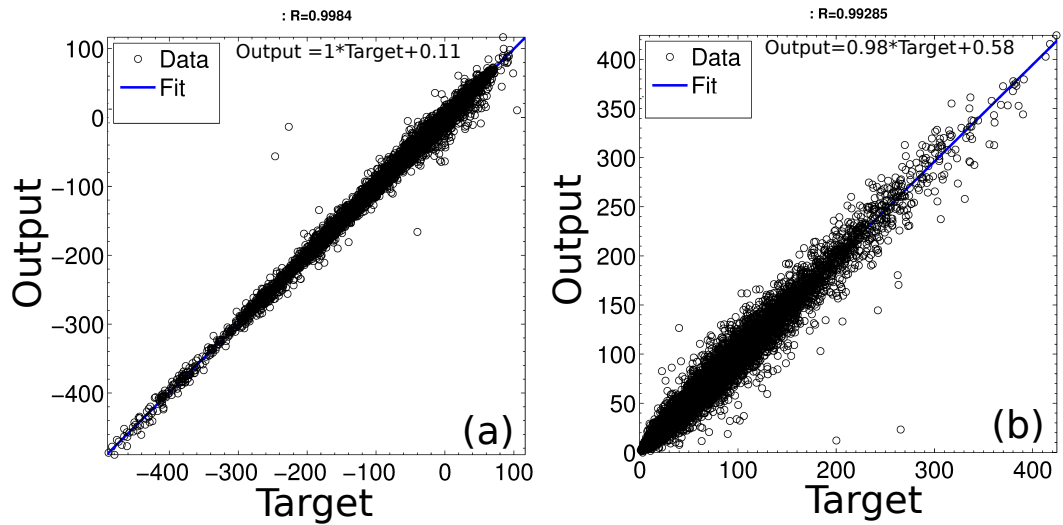}
\end{center}
\caption{The regression of the target and the modeled output by the networks are shown for (a) SYMH and (b) ASYH for training data.}
\label{fig:traireg}
\end{figure}

Figure ~\ref{fig:traireg}a,b shows the linear regression of targets (SYMH/ASYH) and predicated outputs of the networks for best training epochs. The correlation values are almost same $R\sim .99$  for both the networks output-target pairs. However, it is evident that the scatter is better for SYMH as compared to ASYH. The slope value close to unity and low value of intercept of the linear fit between target and output indicate that training is impressive for both the networks.

\subsection{Prediction}
\subsubsection{Geomagnetic storms $-85>$ SYMH $>-210 nT$}

\begin{figure}
\begin{center}
\noindent\includegraphics[width=14cm,angle=0]{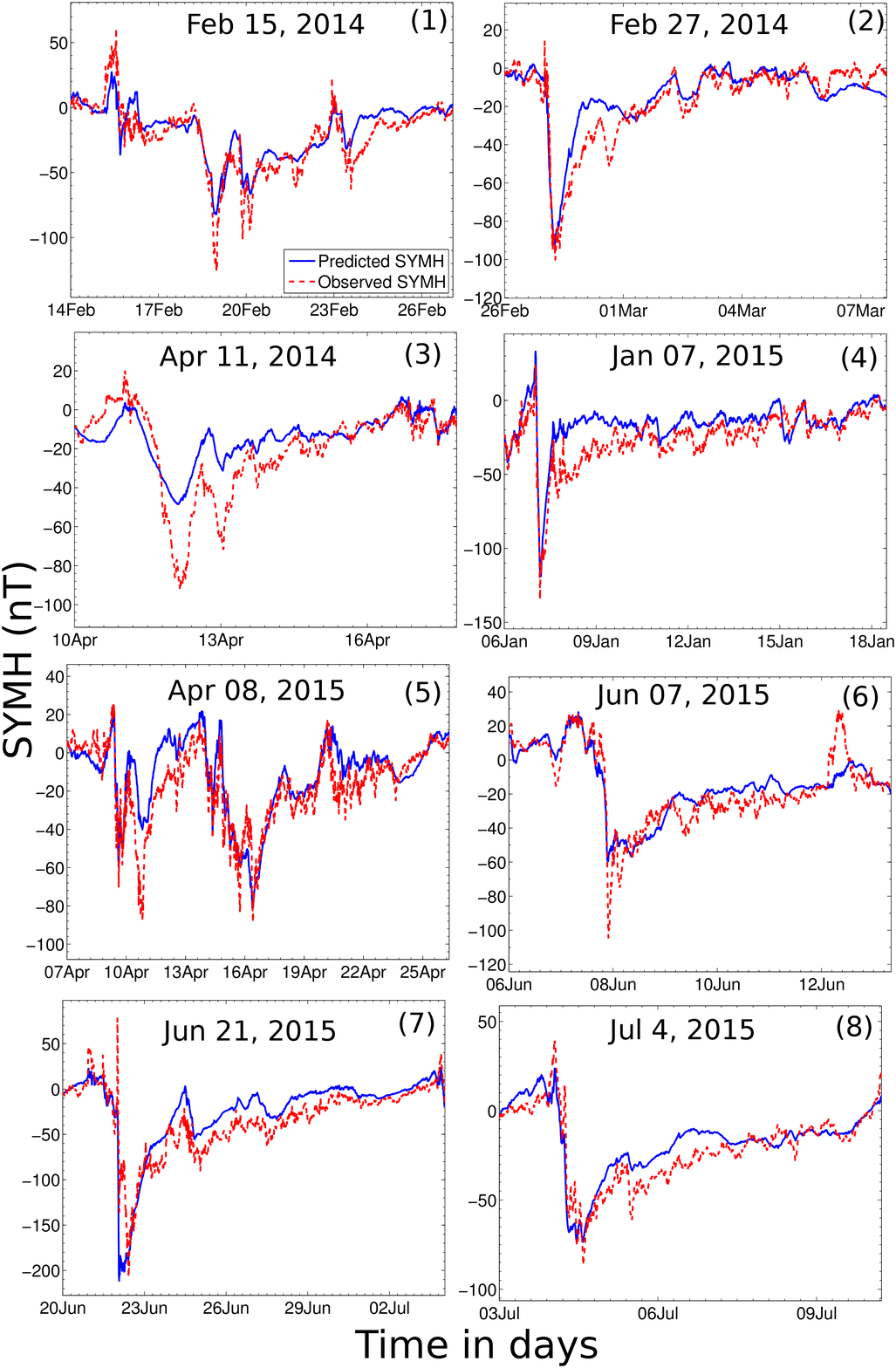}
\end{center}
\caption{Predicated and observed SYMH index for test storms listed in Table ~\ref {table:tab2}. The number indicates the geomagnetic storm number listed in the table. Dotted red curve is observed SYMH and solid blue curve is the predicted SYMH by the network.}
\label{fig:symh1}
\end{figure}

To test the prediction capability of the networks developed here, we have used the geomagnetic storms which were not considered in the training process. The details of these major storms used for prediction are presented in Table \ref{table:tab2}. Figure ~\ref{fig:symh1} shows the predicated (blue line) and observed (dashed red line) values of SYMH for first 8 geomagnetic storms listed in the table. The storm on March 17, 2015 is discussed in detail in the next subsection. In general, all the storms show very good match between predicted and observed SYMH profiles. One can notice that even finer features of timescales 10-30 minutes are reflected in the predicted profiles. 
The prediction of the minimum SYMH matches very well with the observed strength of the storms occurred on Feb 27, 2014; Jan 07, 2015; Jun 21, 2015 and Jul 04, 2015. However, the minimum SYMH of storms occurred on Apr 11, 2014 and Jun 07, 2014 are underestimated by the network. Transient variations like storm sudden commencement (SSC) are reproduced well by the network during storms on Jan 07, 2015 and Jul 04, 2015. Nevertheless, SSC occurred during June 21, 2015 and Feb 27, 2014 are not predicted by the model. Also, two step decrease observed during the main phase of storms (Jun 07, 2015, Jun 21, 2015 and Jul 04, 2015) is reproduced by the network to good extent.

\begin{figure}[!htb]
 \begin{center}
\noindent\includegraphics[width=12cm,angle=0]{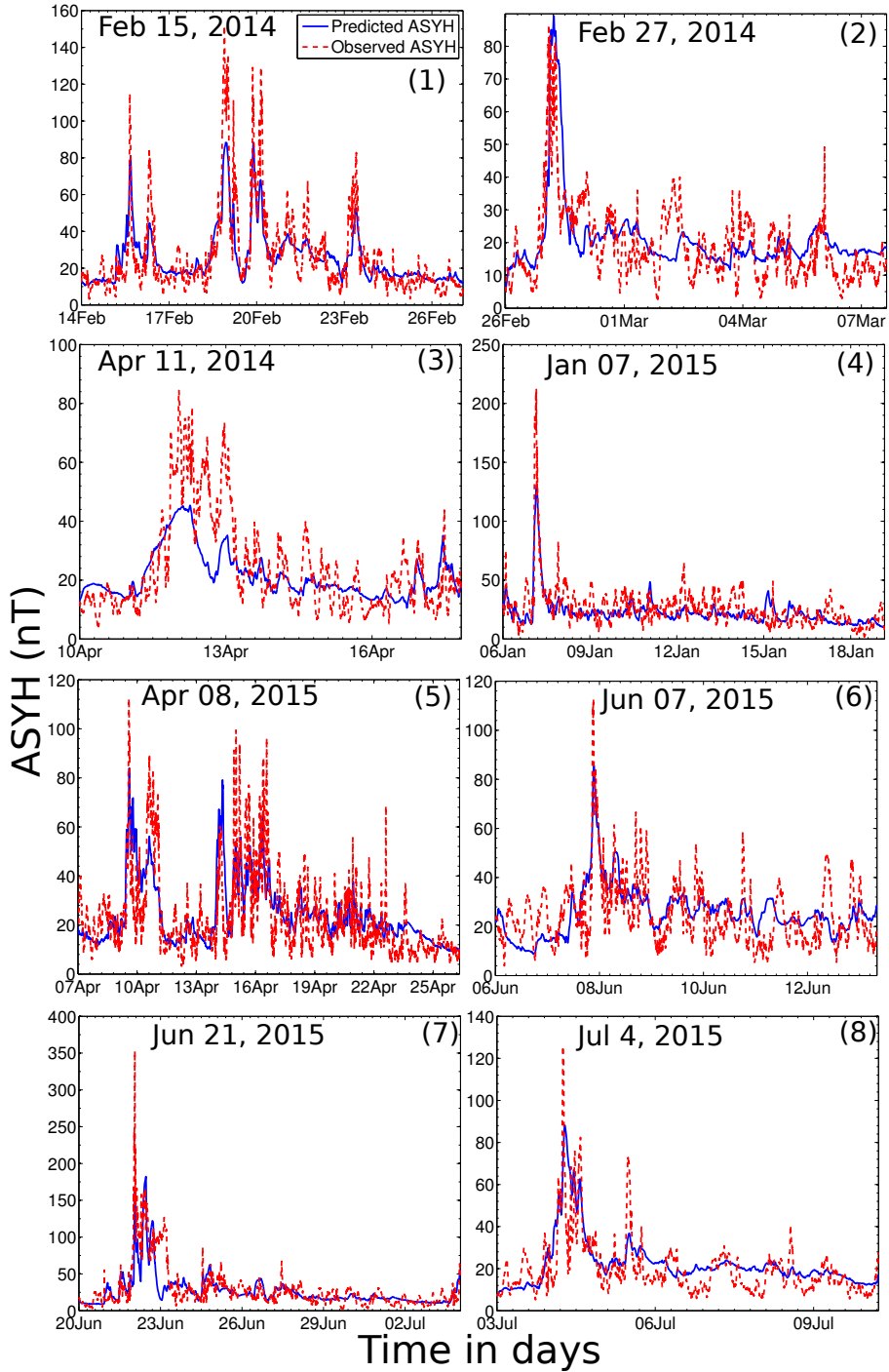}
 \end{center}
 \caption{Predicated and observed ASYH index for test storms listed in Table ~\ref {table:tab2}. The number indicates the geomagnetic storm number listed in the table. Dotted red curve is observed ASYH and solid blue curve is the predicted ASYH by the network.}
 \label{fig:asyh}
 \end{figure}

Figure ~\ref{fig:asyh} shows the predicated (blue line) and observed (dashed red line) values of ASYH index during the storms. The predicted profiles of ASYH match well with the observed profiles. In general, the predication model underestimates the amplitude of  ASYH index. However, the overall temporal profile of ASYH index is well predicated by the network. Note that, finer structures of smaller timescales $\sim 10-30$ minutes are also well mimicked by the model predictions except for Apr 11, 2014 storm. 

\begin{figure}[!htb]
 \begin{center}
\noindent\includegraphics[width=14cm,angle=0]{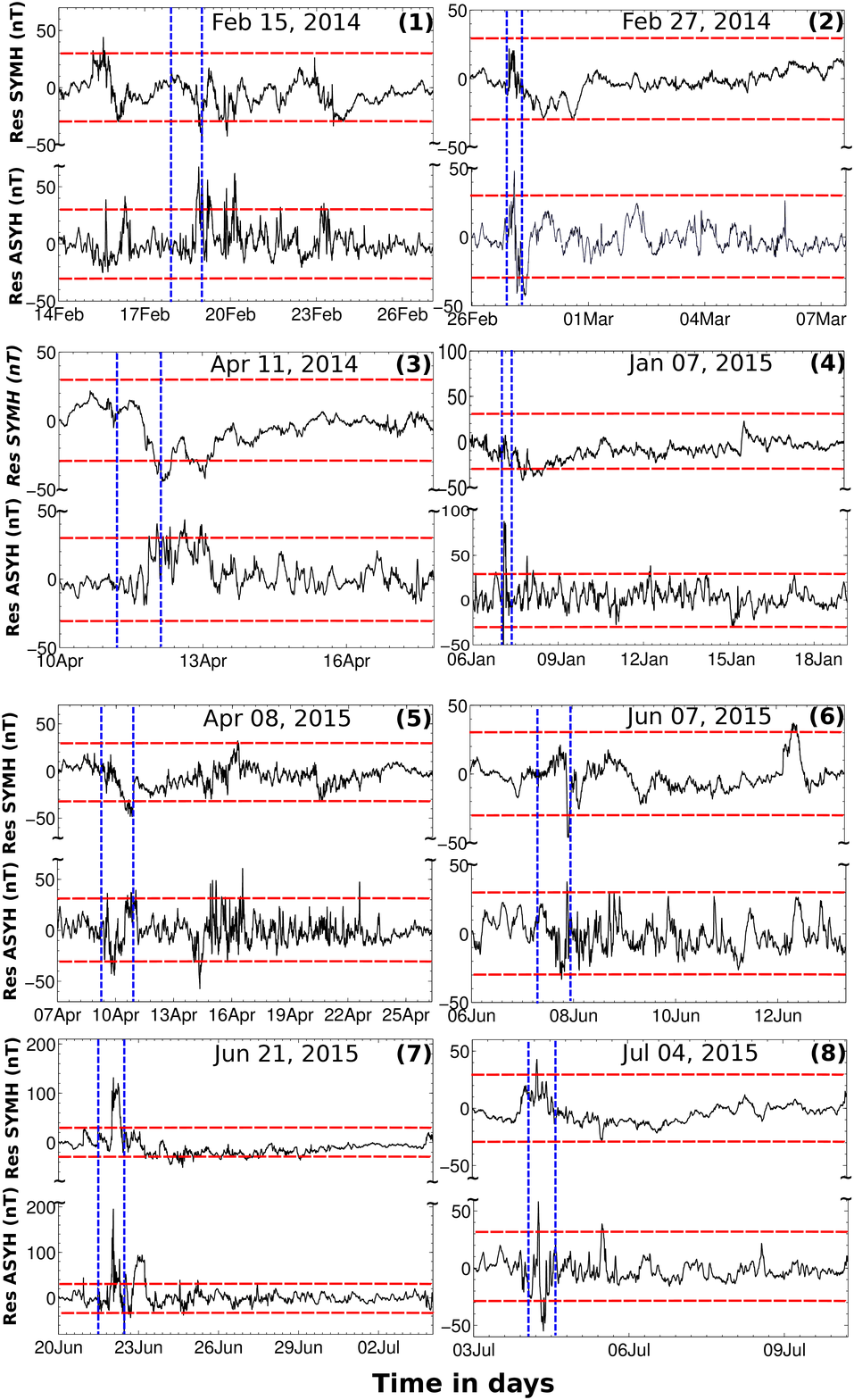}
 \end{center}
 \caption{The residual of predicated and observed values for SYMH and ASYH indices during test storms listed in Table ~\ref {table:tab2}. Dotted horizontal dashed lines mark $\pm 30$ nT noise level (i.e quiet time background). The vertical dashed lines from left represent onset and the end of main phase of the storm respectively.}
 \label{fig:resd}
 \end{figure}

\citet{ tsurutani1997interplanetary} noted a variability of around -30 nT in Dst index, which they considered as a threshold/noise level for geomagnetic storms \citep{munsami2000storm}. Therefore, here we have considered $\pm 30$ nT as the threshold even for SYMH/ASYH, although it is possible that for higher time resolution indices the noise level might be larger.
The residuals of SYMH and ASYH are estimated by subtracting model values from the observed, which are presented in Figure ~\ref{fig:resd} for the selected storms. The noise levels ($\pm 30$ nT) are marked by the red dashed lines  and the main phase bounded by two vertical dashed lines, in each panel. The figure shows that for most of the storms the residuals lie within the noise level, in general. However, for the storm occurred on June 21, 2015 the residuals are well above noise level ($> 100$ nT), in particular during the main phase. Note that the model estimates are based on the interplanetary (external) inputs. Therefore, the higher residual values could be ascribed to the influence of the magnetospheric origin (internal), which is not modeled by the present network. 
Also, one can notice that compared to the residuals in SYMH index, the ASYH residuals are higher in magnitude. This may imply the larger contribution in the ASYH index due to magnetospheric sources, compared to that in SYMH index.

Further, to quantify how good are the networks in predicting these indices, correlation coefficient (R) and RMSE are estimated for the predicted geomagnetic storms which are listed in Table \ref{table:tab2}. A good performance of the networks is more evident from the observed high mean correlation coefficients, $R \sim 0.9$ and $R \sim 0.7$ for SYMH and ASYH indices respectively (see Table \ref{table:tab2}). The table clearly shows the cross correlation is high between predicted and observed SYMH  as compared to correlation between predicated and observed ASYH. This smaller value of the correlation coefficient of ASYH index could be due to the presence of very high frequency fluctuations in ASYH index compared to SYMH index (refer Figure ~\ref{fig:symh1} and \ref{fig:asyh} ). Also, as discussed earlier, ASYH is more complex in nature due to various currents affectingit unlike SYMH.
 
\subsubsection{St. Patrick's day 2015 geomagnetic storm (SYMH $\sim -234 nT$)}
\begin{figure}[!htb]
 \begin{center}
\noindent\includegraphics[width=14cm,angle=0]{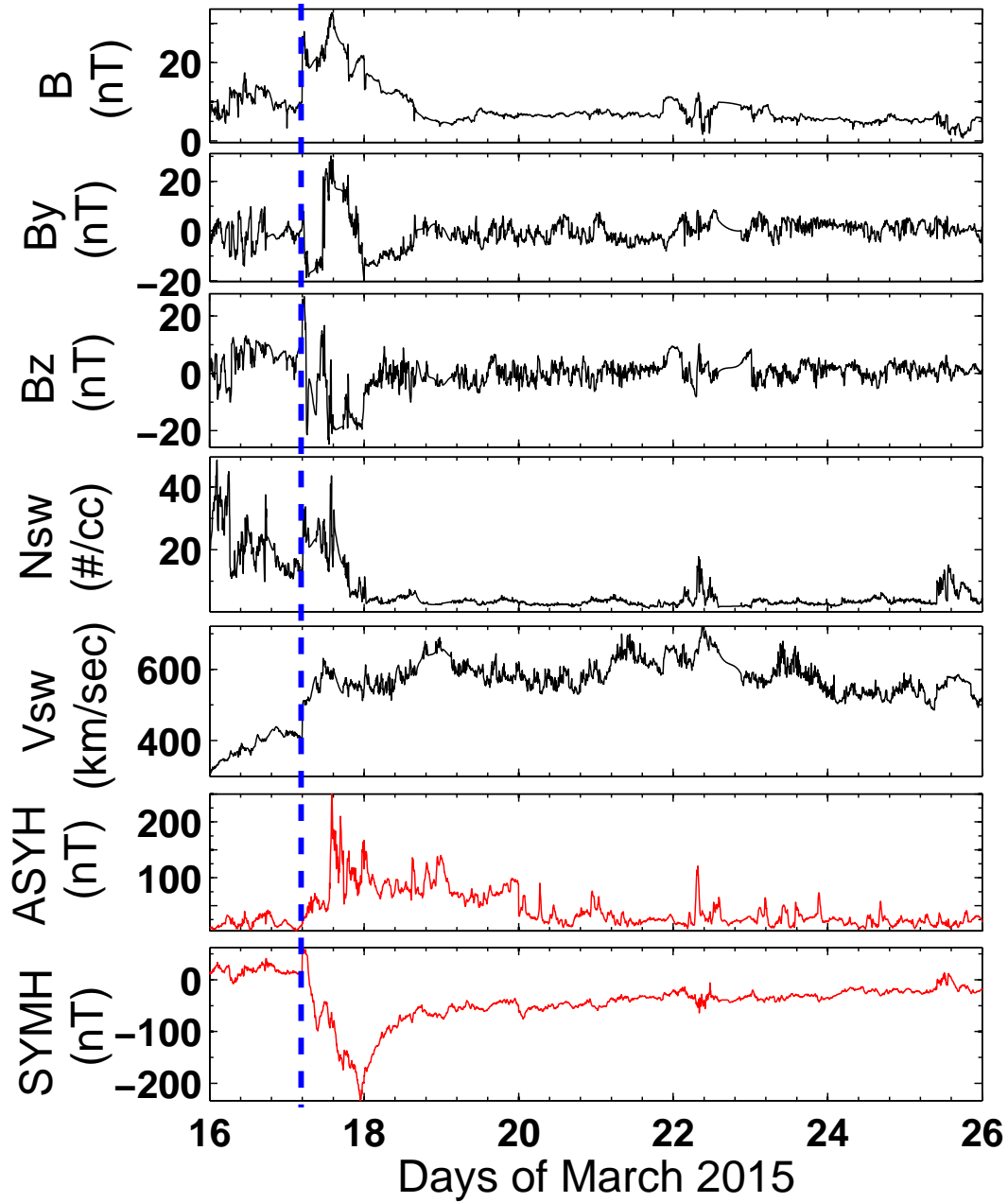}
 \end{center}
 \caption{Interplanetary parameters of intens geomagnetic storm of March 17, 2015: Interplanetary magnetic field (B) and its components By, Bz; solar wind speed (Vsw) and density (Nsw). Geomagnetic indices: ASYH and SYMH. The dashed vertical line marks the onset of the storm}
 \label{fig:IP}
 \end{figure}

\begin{figure}[!htb]
 \begin{center}
\noindent\includegraphics[height=14cm, width=14cm,angle=0]{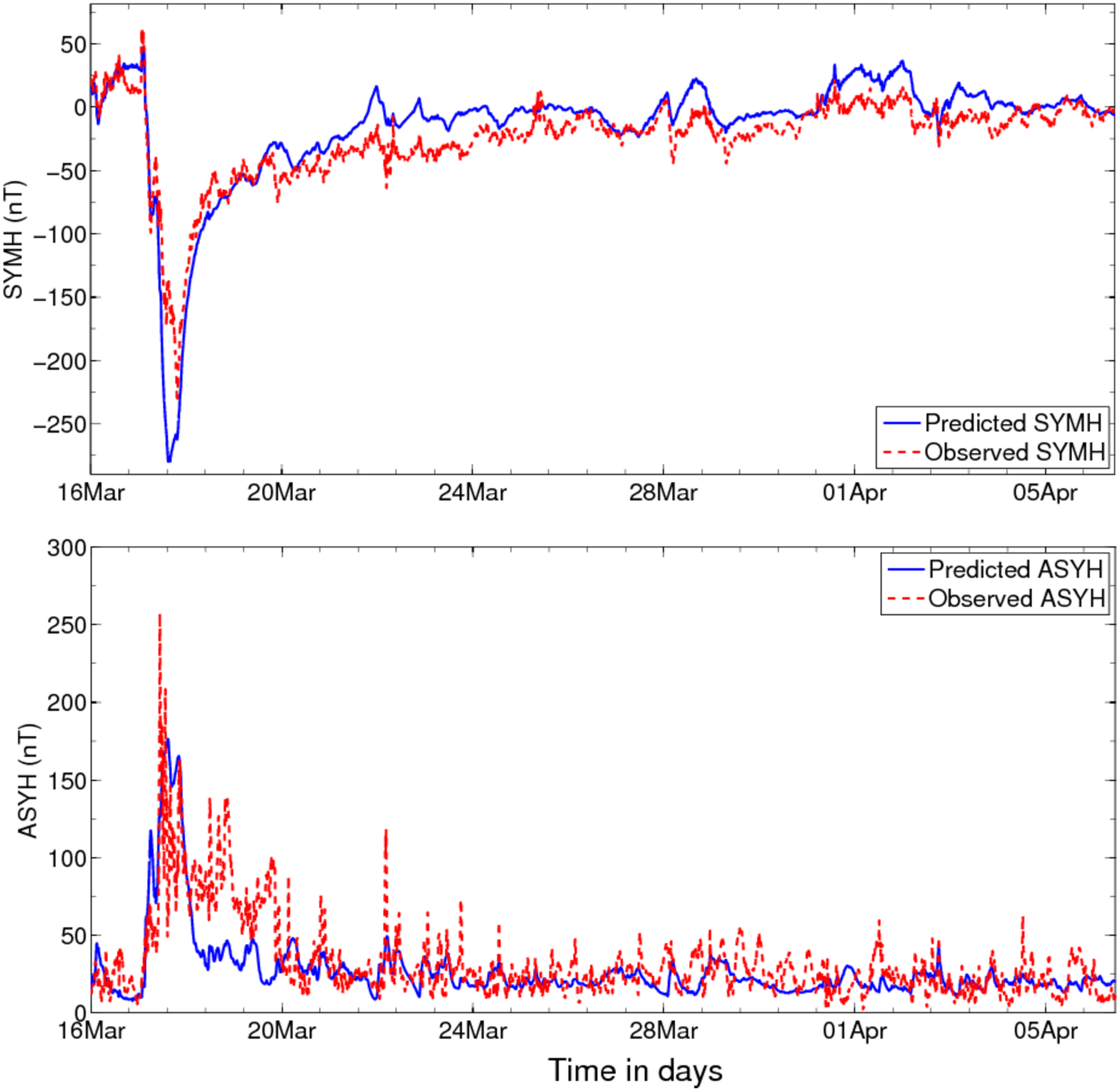}
 \end{center}
 \caption{Predicated and observed (a)SYMH  and (b)ASYH indices for March 17, 2015 geomagnetic storm.}
 \label{fig:storm}
 \end{figure}

Recent storm of $17^{th}$ March 2015  (known as St. Patrick's day 2015 geomagnetic storm) is a great storm of the ongoing $24^{th}$ Solar cycle (see Table \ref{table:tab2}). The parameters of interplanetary disturbance during this storm are presented in Figure ~\ref{fig:IP}. Interplanetary magnetic field (B, By, Bz, nT), Solar wind density (Nsw) and Velocity (Vsw) show the clear enhancement at the onset of the storm on March 17, 2015, time $\sim 05:00$ UT. It is evident that long duration ($\sim 24$ hours) southward Bz with amplitude $\sim 20$ nT has given rise to an intense storm having minimum SYMH $\sim -234$ nT and maximum ASYH of $\sim 250$ nT. The storm shows two distinct steps during main phase, which might be associated with the sudden north-south turnings of the IMF occurred on March 17, 2015 at $\sim 11 $UT. ASYH index is enhanced during the main and early recovery phase of the storm. The ASYH index is almost of the same magnitude that of SYMH during the main phase. The prediction results obtained from the present models along with the observed indices are displayed in Figure ~\ref{fig:storm}. The trained NARX network for SYMH predicts the observed SYMH very well including the sudden storm commencement, two steps in the main phase and small transient fluctuations during the recovery phase of the storm. The predicted storm time minimum SYMH is close to the observed one. The correlation between predicated and observed SYMH is very high ($R\sim 0.93$) and RMSE value is low ($\sim 21$ nT). The match between predicted and observed ASYH is excellent during the main phase. For ASYH, correlation between predicated and observed ASYH is $R\sim 0.7$ and RMSE value is $\sim 20$ nT.

\section{Discussion and conclusions}
As ASYH index is of paramount importance to unravel the information about the asymmetric response of the magnetosphere especially during geomagnetic storms, present study attempts to predict ASYH index for the first time. We have applied NARX neural network to 92 geomagnetic storms occurred between 1998-2013. The developed networks successfully predict SYMH and ASYH indices about an hour prior to the onset of storm provided the real time upstream solar wind data is available. The need of 30 min input history and 120 minute feedback for better predictions imply the role of preconditioning of magnetosphere i.e the future state of contributing currents in indices depends on their present and past values. We have examined the prediction for 9 geomagnetic storms from solar cycle 24, occurred during January, 2014-July, 2015. These storms include the major storm occurred on St. Patrick's day, 2015, which is the most intense storm occurred so far in solar cycle 24.

The ability of NARX having feedback from output enabled us to model these indices quite accurately. The temporal variations of the order of 10-30 minute are well predicted by both the networks. The network trained for SYMH index predicts SYMH very well and observed average correlation between predicted and observe SYMH is high (R $\sim 0.88$) i.e almost $\sim 77\%$ variations of SYMH are modeled by the network. The average RMSE is about 13.98 nT and matches with the observations by \citep{cai2009prediction}. Therefore, the prediction performance of the network is almost same as that of ANN constructed by \citet{cai2009prediction}.
However, as noted earlier the prediction accuracy varies from storm to storm. \cite{munsami2000storm} also observed mismatch between predicated and observed Dst index, which they thought to be due to other than external drivers such as substorms. However they did not observe noticeable improvement in Dst prediction even by considering inputs from substorm activity. This could be due to the contribution to Dst from other processes such as wave-particle interaction, charge exchange, ionospheric outflow of $O^{+}$ ions, particle loss to atmosphere and magnetopause \citep{Daglis1999, liemohn2001dominant}. The residuals between observed and predicted values of SYMH generally lie within the noise level of $\pm 30$ nT. However, note that sometimes high residual values are  observed above the noise level, especially during main phase of geomagnetic storms.

In general, the prediction of ASYH index is very good, within the noise level of $\pm 30$ nT.  More than $50 \%$ variations of ASYH are explained by the present network. This implies that the variation in asymmetric ring current could be explained by solar wind parameters. However, during the main and early recovery phase of storms, the residuals (observed-modeled) are above noise level, which could be ascribed to the internal magnetospheric processes such as field aligned currents, particle loss.



The present study demonstrates that developed networks are capable in predicting SYMH and ASYH indices and hence can be implemented for the real time forecasting. Interestingly, even though ASHY is a good proxy for internal variability of asymmetric ring current ANN could model large part of the variations using external (solar wind) parameters. The reliable forecast of SYMH and ASHY indices will help space weather community and space programs to get early information on the strength of geomagnetic disturbances and their asymmetric geomagnetic response. As this is the first attempt to predict ASYH using ANN, in future the prediction may be improved by considering inputs representative of internal magnetospheric dynamics.

\begin{acknowledgments}
The solar wind parameters, interplanetary magnetic field, and geomagnetic indices used in this paper are obtained from CDAWEB (\url{http://cdaweb.gsfc.nasa.gov/}). We thank ACE Science Center for making data available in public domain. 
\end{acknowledgments}

%
%
%
%
%
%
%
%
%

\bibliographystyle{agufull08} 
\bibliography{ref}

\clearpage

\clearpage

\clearpage

\clearpage

\clearpage

\clearpage

\clearpage

\clearpage
%

\clearpage
\begin{table}
\scriptsize

\centering
\caption{Geomagnetic storm durations of 2006-2013 considered for training the networks in addition to 67  geomagnetic storms of \citet{cai2009prediction}}
\label{table:tab1}
\begin{tabular}{|c| c c c|c c c |c|}
\hline
\multicolumn{1}{|c|}{\textbf{}}    & \multicolumn{3}{c|}{\textbf{   Start}}                                                                          & \multicolumn{3}{c|}{\textbf{   End}}                                                                            & \multicolumn{1}{l|}{\textbf{}}                \\ \hline
\multicolumn{1}{|l|}{\textbf{No.}} & \multicolumn{1}{l}{\textbf{Year}} & \multicolumn{1}{l}{\textbf{Month}} & \multicolumn{1}{l|}{\textbf{Day}} & \multicolumn{1}{l}{\textbf{Year}} & \multicolumn{1}{l}{\textbf{Month}} & \multicolumn{1}{l|}{\textbf{Day}} & \multicolumn{1}{l|}{\textbf{Min. SYMH (nT)}} \\ \hline
1 & 2006 & 4  & 3  & 2006 & 4  & 8  & -93  \\
2 & 2006 & 4  & 8  & 2006 & 4  & 13 & -107 \\
3 & 2006 & 4  & 13 & 2006 & 4  & 21 & -111 \\
4 & 2006 & 8  & 18 & 2006 & 8  & 26 & -95  \\
5 & 2006 & 12 & 14 & 2006 & 12 & 18 & -211 \\
6 & 2008 & 3  & 7  & 2008 & 3  & 19 & -100 \\
7 & 2009 & 7  & 19 & 2009 & 7  & 30 & -95  \\
8 & 2011 & 3  & 9  & 2011 & 3  & 17 & -92  \\
9 & 2011 & 5  & 26 & 2011 & 6  & 17 & -94  \\
10 & 2011 & 8  & 5  & 2011 & 8  & 23 & -126 \\
11 & 2011 & 9  & 26 & 2011 & 10 & 14 & -116 \\
12 & 2011 & 10 & 24 & 2011 & 11 & 18 & -160 \\
13 & 2012 & 1  & 21 & 2012 & 2  & 12 & -88  \\
14 & 2012 & 3  & 6  & 2012 & 3  & 26 & -150 \\
15 & 2012 & 4  & 23 & 2012 & 4  & 30 & -125 \\
16 & 2012 & 7  & 14 & 2012 & 7  & 23 & -123 \\
17 & 2012 & 9  & 28 & 2012 & 10 & 5  & -138 \\
18 & 2012 & 10 & 7  & 2012 & 10 & 11 & -116 \\
19 & 2012 & 10 & 11 & 2012 & 10 & 23 & -106 \\
20 & 2012 & 11 & 12 & 2012 & 11 & 17 & -118 \\
21 & 2013 & 3  & 17 & 2013 & 3  & 26 & -132 \\
22 & 2013 & 5  & 31 & 2013 & 6  & 6  & -137 \\
23 & 2013 & 6  & 6  & 2013 & 6  & 9  & -88  \\
24 & 2013 & 6  & 27 & 2013 & 7  & 4  & -111 \\
25 & 2013 & 10 & 1  & 2013 & 10 & 8  & -90  \\ 
\hline
\end{tabular}
\end{table}

\clearpage
\begin{landscape}
\begin{table}
\scriptsize
\centering
\caption{Geomagnetic storm durations considered for testing the networks of SYMH and ASYH}
\label{table:tab2}
\begin{tabular}{|c| c c c|c c c |c|c c|c c|}
\hline
\multicolumn{1}{|c|}{\textbf{}}    & \multicolumn{3}{c|}{\textbf{   Start}}                                                                          & \multicolumn{3}{c|}{\textbf{   End}} & \multicolumn{1}{|c|}{\textbf{Strength}} & \multicolumn{2}{c|}{\textbf{SYMH}} & \multicolumn{2}{c|}{\textbf{ASYH}}                                                                                              \\ 
\hline
\multicolumn{1}{|l|}{\textbf{No.}} & \multicolumn{1}{l}{\textbf{Year}} & \multicolumn{1}{l}{\textbf{Month}} & \multicolumn{1}{l|}{\textbf{Day}} & \multicolumn{1}{l}{\textbf{Year}} & \multicolumn{1}{l}{\textbf{Month}} & \multicolumn{1}{l|}{\textbf{Day}} & \multicolumn{1}{l|}{\textbf{Min. SYMH (nT)}} & \multicolumn{1}{l}{\textbf{R}} & \multicolumn{1}{l|}{\textbf{RMSE}} & \multicolumn{1}{l}{\textbf{R}} & \multicolumn{1}{l|}{\textbf{RMSE}} \\ 
\hline
1 & 2014 & 2 & 15 & 2014 & 2 & 26 & -127 & 0.9  & 12.51 & 0.85 & 12.51 \\
2 & 2014 & 2 & 27 & 2014 & 3 & 7  & -101 & 0.9  & 9.17  & 0.66 & 9.84  \\
3 & 2014 & 4 & 11 & 2014 & 4 & 17 & -92  & 0.87 & 14.55 & 0.78 & 11.54 \\
4 & 2015 & 1 & 7  & 2015 & 1 & 18 & -135 & 0.84 & 13.44 & 0.76 & 12.59 \\
5 & 2015 & 4 & 8  & 2015 & 4 & 25 & -89  & 0.89 & 10.3  & 0.69 & 12.05 \\
6 & 2015 & 6 & 7  & 2015 & 6 & 12 & -105 & 0.85 & 12.44 & 0.58 & 11.17 \\
7 & 2015 & 6 & 21 & 2015 & 7 & 3  & -208 & 0.87 & 22.58 & 0.76 & 20.25 \\
8 & 2015 & 7 & 4  & 2015 & 7 & 10 & -87  & 0.91 & 9.57  & 0.73 & 10.07 \\
9 & 2015 & 3 & 17 & 2015 & 4 & 5  & -234 & 0.93 & 21.34 & 0.68 & 20.43 \\
\hline
\end{tabular}
\end{table}
\end{landscape}

\bibliographystyle{agufull08}
\bibliography{ref}

\end{article}

 %
%
%
%
%


\end{document}